\documentclass[aps,prd,twocolumn,superscriptaddress,preprintnumbers,floatfix,nofootinbib]{revtex4-1}
\usepackage{multirow}
\usepackage{graphicx}
\usepackage{dcolumn}
\usepackage{bm}
\usepackage{hyperref}
\usepackage{float}
\usepackage{color}
\usepackage[german=quotes]{csquotes}

\newcommand{\be}{\begin{equation}}
\newcommand{\ee}{\end{equation}}
\newcommand{\bea}{\begin{eqnarray}}
\newcommand{\eea}{\end{eqnarray}}

\usepackage{mathrsfs}
\usepackage{amsmath, amsthm, amssymb}
\usepackage{color}
\usepackage{epstopdf}
\usepackage{amssymb}
\usepackage{verbatim}
\usepackage{hyperref}
\usepackage{enumerate}
\usepackage{float}
\usepackage[caption = false]{subfig}
\usepackage[normalem]{ulem}  
\usepackage{epsfig}

\newcommand{\physrep}{Phys.~Rep.}   

\begin{document}

\title{Vector Dark Matter Halo: From Polarization Dynamics to Direct Detection}

\author{Jiajun Chen}
\email{chenjiajun@swu.edu.cn}
\thanks{Corresponding author}

\affiliation{School of Physical Science and Technology, Southwest University, Chongqing 400715, China}
\affiliation{Institut f\"ur Experimentalphysik, Universit\"at Hamburg, Hamburg, Germany.}

\author{Le Hoang Nguyen}
\email{le.hoang.nguyen@uni-hamburg.de}
\affiliation{Institut f\"ur Experimentalphysik, Universit\"at Hamburg, Hamburg, Germany.}

\author{David J. E. Marsh}
\email{david.j.marsh@kcl.ac.uk}
\affiliation{Theoretical Particle Physics and Cosmology, King's College London, Strand, London, WC2R 2LS, United Kingdom}

\begin{abstract}
This study investigates the characteristic polarization formation and evolution of vector dark matter (VDM) in the outer halo of galaxies. By employing numerical simulations, we analyze the behavior of VDM under different initial conditions—homogeneous, isotropic, and partially polarized. The simulations solve the Schrödinger-Poisson equations, examining the spin density distribution and its evolution during gravitational collapse and halo formation. Our results reveal that VDM forms halos and central Proca stars from homogeneous and isotropic conditions, with the polarization density fluctuation amplitude mirroring VDM matter density. In scenarios with no initial polarization, spin density remains stable in the halo core but fluctuates in outer regions. Partially polarized initial conditions lead to a conservation of total polarization, with increased core polarization resulting in opposite polarization in the periphery. We examine the novel consequences of the partially polarized state for direct detection of dark photons, i.e. VDM kinetically mixed with ordinary photons.
\end{abstract}

\maketitle

\section{Introduction}
The composition of dark matter (DM) remains a central unresolved issue in modern cosmology. Observations find that dark matter composes  about $27\%$ of the total mass-energy of the universe~\cite{Aghanim:2018eyx}. A popular hypothesis is that dark matter is composed of light scalar or pseudoscalar fields, with QCD axions and similar axion-like particles being typical examples. These models are collectively referred to as ``Scalar Dark Matter" (SDM), see e.g. Refs.~\cite{Dine:1982ah,Preskill:1982cy,Abbott:1982af,Suarez:2013iw,Arias:2012az,Marsh:2015xka,OHare:2024nmr}.

A related model considers how spin-1 particles may also acquire small mass, making up light bosonic dark matter. This model is commonly referred to as ``hidden" or ``dark photon" dark matter, which we generically term ``Vector Dark Matter" (VDM)~\cite{Graham:2015rva}. VDM has gained attention in recent years due to the discovery of feasible production mechanisms similar to scalar vacuum realignment. 

Refs.~\cite{Arias:2012az, Horns:2012jf} laid out a framework for haloscope experiments searching for VDM, showing how one can interpret the results of axion haloscopes with the $B$ field turned off in terms of limits on VDM-photon mixing. Meanwhile, Ref.~\cite{Caputo:2021eaa} discussed the scenario where VDM haloscopes are subjected to the sidereal modulation of the signal output power due to the alignment of the spin-vector of the local VDM with the sensitive polarizations of the detector. Thus, one distinguishes between two main possibilities for the VDM polarization: \emph{fixed spin-vector}, or \emph{random spin-vector}. The case of fixed spin-vector is the most conservative since the (unknown) fixed value could be misaligned with the optimal sensitive polarization of the experiment. 

VDM is described by a four vector field $A_\mu$. Given a mass, $m$, this leads to three polarization states: two transverse helicities, $A^\pm_T$, related to circular polarization of light, and one longitudinal mode, $A_L$, related to the mass. Most production mechanisms for VDM~\cite{Graham:2015rva,Agrawal:2018vin,Dror:2018pdh,Long:2019lwl} lead to either the longitudinal mode, or the transverse mode, or both, being populated. As described in detail below and in Refs.~\cite{Zhang:2021xxa,Amin:2022pzv,Gorghetto:2022sue,Adshead:2021kvl,Jain:2021pnk,Gosenca:2023yjc,Glennon:2023jsp, PhysRevD.108.083021}, in the non-relativistic limit appropriate for DM in our galaxy VDM is described by three Schr\"{o}dinger fields $\psi_i$ aligned with the Cartesian coordinate axes and composed into a vector $\bm{\Psi}$. {The fields $\psi_i$ give the amplitude of the vector potential $A_i$ in the non-relativistic limit (WKB approximation) in which the ``fast'' Compton timescale, $1/m$ is factored out. The $\psi_i$ fields interact with one another via a non-linear Schr\"odinger equation, with the potential determined via Poisson's equation for the mass density of VDM, which is proportional to $|\bm{\Psi}|^2$. This is in exact analogy with the case of scalar/axion DM (see e.g. Ref.~\cite{Marsh:2015daa} for a derivation of the Schr\"odinger-Poisson equation from the relativistic Klein-Gordon-Einstein equations). It is important to note that the relevant description arises from the \emph{classical} field equations: the appearance of the Schr\"odinger equation is a mathematical coincidence and is not the quantum equation of the same name.}

Circular/elliptical polarization is now encoded in the spin angular momentum density $\bm{s}= i \bm{\Psi}\times \bm{\Psi}^*$. Initial conditions of either longitudinal or transverse polarizations in terms of $A_\mu$ in general populate all $\psi_i$ modes, and thus give rise a random distribution for the spin polarization (see e.g. Ref.~\cite{Gorghetto:2022sue,Jain:2023ojg,Amaral:2024tjg}). Even in such a random model, however, the polarization will (just like scalar/axion DM~\cite{Marsh:2015xka,OHare:2024nmr}) display a coherence length and time governed by gravitational dynamics. On the other hand, there also exist Bianchi universe and Lorentz violating models that can preferentially populate given $\psi_i$ modes and give rise to directionally polarized states~\cite{Arias:2012az,Alonso-Alvarez:2019ixv} (although see Refs.~\cite{Himmetoglu:2008zp,Himmetoglu:2009qi,Karciauskas:2010as}).

Ref.~\cite{Caputo:2021eaa} noted that, at the time of publication, the precise spin-vector distributions after accounting for non-linear structure formation had not been considered. It is the goal of the present work to study exactly this effect. Previous works have studied the collapse and condensation of vector solitons and DM halos~\cite{PhysRevD.108.083021,Jain:2023ojg,Gorghetto:2022sue,Amin:2022pzv}. It has been found that non-linear gravitational collapse leads to a kind of thermalisation and equipartition between the spin polarization modes~\cite{Amaral:2024tjg}. Ref.~\cite{Amaral:2024tjg} considered how the polarization evolves on timescales of order the coherence time, $\tau_c=1/mv_{\rm vir}^2$ with $v_{\rm vir}$ the halo virial velocity, finding that the polarization follows the density coherence and that the vector field sweeps out ellipses in real space. 

The coherence scale dynamics of DM is relevant to laboratory searches operating at the very lowest masses/frequencies of $m\lesssim 10^{-15}\text{ eV}$, $\nu\lesssim 1\text{ Hz}$ (and such dynamics may provide insight into the quantum state of DM~\cite{Marsh:2022gnf,Eberhardt:2023axk}). In this work, we are concerned with the spin polarization distribution on time scales much larger than the coherence time and throughout the DM halo, since we have in mind application to dish antenna VDM searches such as BRASS-p~\cite{Nguyen:2023uis}, which have polarization sensitivity in the $\nu\gtrsim 10\text{ GHz}$ frequency range with $m\gtrsim 40 \,\mu\text{eV}$.

We consider a series of different simulations with different initial spin states for VDM. Our simulations are not cosmological, but instead they condense VDM halos from homogeneous an isotropic noise, following Refs.~\cite{Levkov:2018kau,PhysRevD.104.083022,PhysRevD.108.083021,Jain:2023ojg,Chen:2021nnf}. The statistical properties of the DM in such halos are known to be similar to those formed cosmologically (e.g. Ref.~\cite{Schive:2014dra,Schive:2014hza}), but can be simulated at lower computational cost. We consider models with random spin polarization, i.e. random mixes between transverse and longitudinal polarizations. We also consider, for the first time, a novel mixed state polarization which mixes the fixed and random models. We do not simulate fully polarized cases (either spin or linear), as these have trivial evolution in terms of polarization.

We start in Sec.~\ref{sec:SP_IC} by introducing the Schr\"{o}dinger-Poisson (SP) equations for non-relativistic VDM and our initial distributions. Sec.~\ref{sec:Evo_number_spin_density} studies how VDM's particle number and spin density evolve thorughout the halo under different initial conditions using numerical simulations.
Sec.~\ref{sec:spin_density_and_mixing state} presents a physical interpretation and model for the novel mixed polarization state. Sec.~\ref{sec:direct_searches} presents the consequences of this mixed state for direct detection experiments. We present our conclusions in Sec.~\ref{sec:conclusion}.

\section{Equations of Motion and Initial Conditions}
\label{sec:SP_IC}
A real vector field, $A_\mu$, with mass $m$, minimally coupled to gravity, has the following action:
\begin{equation}
S =\int d^4 x \sqrt{-g}\left( -\frac{1}{4}F_{\mu\nu}F^{\mu\nu}-\frac{1}{2}m^2 A_\mu A^\mu\right)\, ,
\label{eq:action}
\end{equation}
where $F_{\mu\nu}=\partial_\mu A_\nu-\partial_\nu A_\mu$ is the field strength tensor, and indices are raised and lowered by the metric $g_{\mu\nu}$ with determinant $g$. The equation of motion is:
\begin{equation}
\nabla_\mu F^{\mu\nu}= m^2 A^\nu\, ,
\label{eq:proca_eq}
\end{equation}
known as the Proca equation, which implies the Lorentz condition, $\nabla_\mu A^\mu=0$, as a constraint, thus reducing the number of degrees of freedom of $A_\mu$ to three dynamical fields (see e.g. Refs.~\cite{Brito:2015pxa,Gorghetto:2022sue}). We consider the non-relativistic limit, where these degrees of freedom are written as:
\begin{equation}
A_i = \frac{1}{\sqrt{2m}}\left( \psi_i e^{imt}+\psi_i^* e^{-imt}\right)\, .
\label{eq:A_i}
\end{equation}

In this limit, in exact analogy to a real scalar field (see e.g. Ref.~\cite{Marsh:2015xka}), for small perturbations of the metric about Minkowski space by the Newtonian potential $V$, and at lowest order in the non-relativistic limit the Einstein-Proca equations reduce to three copies of the SP equations for each complex field $\psi_i$:
\begin{eqnarray}
i\frac{\partial}{\partial{t}} \bm{\Psi}&=&-\frac{1}{2m}\nabla^2\bm{\Psi} + m V\bm{\Psi},
\label{eq:SP1}
\\
\nabla^2{V}&=&4 \pi G m\left(\bm{\Psi}^{\dag}\bm{\Psi}-n\right),
\label{eq:SP2}
\end{eqnarray}
where $\bm{\Psi}=(\psi_1,\psi_2,\psi_3)^T$ is the wave function vector, $n=n_1+n_2+n_3$ is the mean total number density, $G$ is Newton's gravitational constant, and natural units, $\hbar=c=1$, are used. The number density, mass density, and spin density are
$N = \bm{\Psi}^*\bm{\Psi}$, $\rho =  m\bm{\Psi}^*\bm{\Psi}$, $\bm{s} = \left(i \bm{\Psi} \times \bm{\Psi}^{*}\right)$.
The spin density $\bm{s}$ arises from the $\mathrm{SO}(3)$ symmetry of $\bm{\Psi} = (\psi_1, \psi_2, \psi_3)^T$, describing the spatial distribution of the conserved angular momentum.
The $SO(3)$ symmetry leads to the conservation of intrinsic spin angular momentum, $\bm{S}$:

\begin{eqnarray}
\bm{S} &=& \int d^3 x\, \left(i \bm{\Psi} \times \bm{\Psi}^{*}\right).
\label{eq:conser_S}
\end{eqnarray}

Introducing the dimensionless quantities
\begin{eqnarray}
    x=\widetilde{x}/(m v_0),\quad t=\widetilde{t}/(mv_0^2),\quad V=\widetilde{V} v_0^2,\nonumber\\
    \psi_i=\widetilde{\psi_i}v_0^2\sqrt{m/(4 \pi G)},
    \label{eq:dim}
\end{eqnarray}
where $v_0$ is a reference velocity (e.g. the characteristic velocity of the initial state), we obtain the dimensionless equations that we will solve numerically
\begin{eqnarray}
    i\frac{\partial}{\partial{\widetilde{t}}}\widetilde{\bm{\Psi}}&=&-\frac{1}{2}\widetilde{\nabla}^2\widetilde{\bm{\Psi}} + \widetilde{V}\widetilde{\bm{\Psi}},
    \label{eq:SP1_dim}\\
    \widetilde{\nabla}^2{\widetilde{V}}&=&\widetilde{\bm{\Psi}}^{\dag}\widetilde{\bm{\Psi}}-{\widetilde{n}}.
    \label{eq:SP2_dim}
\end{eqnarray}

Similarly to previous studies on the condensation of VDM~\cite{PhysRevD.108.083021,Jain:2023ojg}, we generate homogeneous and isotropic initial conditions in a periodic box of size $L$. Each component of the vector field is assumed to be a Gaussian random field with a momentum distribution of $|\psi_i(\bm{p})|^2 \propto \delta(|\bm{p}|-m v_0)$. Specifically, given the momentum distribution we perform an inverse Fourier transform on $\psi_i(\bm{p}) e^{i A}$ with $A$ a random phase uniformly drawn from $[0,2\pi)$ for each non-zero $\psi_i(\bm{p})$. The wave function is normalized to give a specified total number of non-relativistic bosons for the $i-$th component in the box, $N_i\equiv n_i L^3$. The initial condition is then evolved by solving the SP equations numerically using a fourth-order time-splitting pseudospectral method \cite{Du:2018qor,Mocz:2017wlg,PhysRevD.108.083021,PhysRevD.104.083022,PhysRevD.106.023009,Chen:2024pyr}.

\begin{figure*}[htbp]
\centering
\textbf{}\par\medskip
\begin{minipage}{0.49\textwidth}
  \centering
  \vspace{2mm} 
  \includegraphics[width=\linewidth]{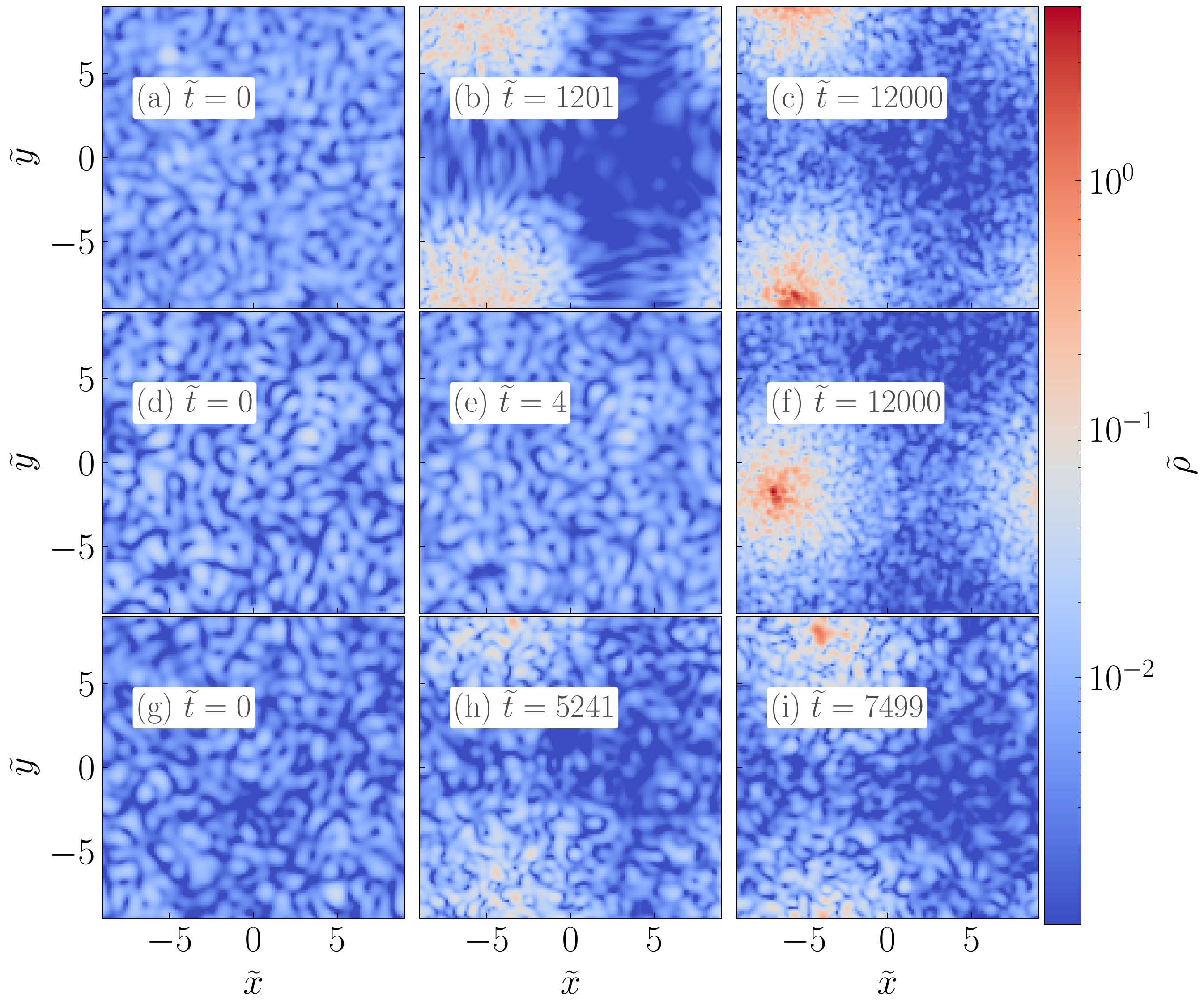}
\end{minipage}\hfill
\begin{minipage}{0.49\textwidth}
  \centering
  \includegraphics[width=\linewidth]{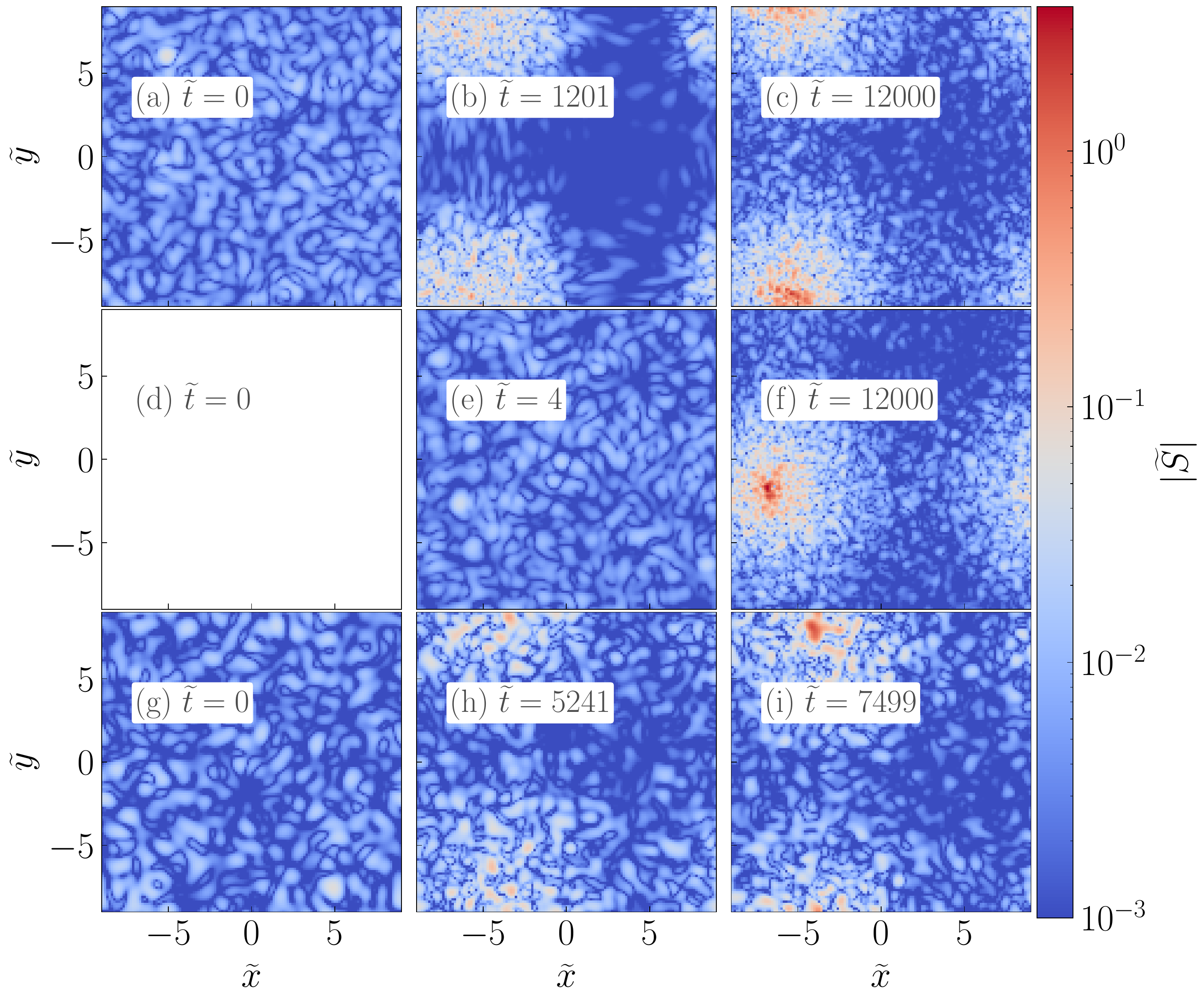}
\end{minipage}
\caption{Slice of particle number density (left figure) and spin density (right figure) evolution for the three initial conditions within the simulation box of $\widetilde{L} = 64$. In each figure, top row indicates the formation of a halo and Proca star from homogeneous and isotropic CF initial conditions with $\widetilde{N}_1=\widetilde{N}_2 =\widetilde{N}_3 \approx 630$. The middle row is the evolution for the RF initial condition (exact zero spin density). 
    The bottom row is for the CF/$\psi_i = 0$ correlated initial condition with partial polarization. Here we take $\widetilde{\psi_x} = \psi_1$, $\widetilde{\psi_y} = (2\psi_2 -i\psi_1)/\sqrt{5}$, $\widetilde{\psi}_z = 0$ with $\widetilde{N}_1=\widetilde{N}_2 \approx 630$, $\widetilde{N}_3 = 0$. The figures on the right, from top to bottom, respectively refer to the slice of spin density.}
\label{fig:Slice_Rho_spin}
\end{figure*}

We consider three types of initial condition: 
\begin{enumerate}
    \item CF -- three independent \textit{complex fields}. {We first use the initial conditions employed in the study of Proca star condensation time~\cite{PhysRevD.108.083021}.} In this case, a random polarization value occurs throughout the box at the initial time. 
    \item RF -- three independent \textit{real fields}. {This initial condition has zero spin density, and maintains it exactly due to symmetry. It serves as a test of our numerical and physical understanding.}
    \item Correlated initial conditions, (labeled {CF/$\psi_i = 0$}). {In addition to the initial conditions with independent components, we also consider those} specified by the correlation matrix $C_{ij}$~\cite{PhysRevD.108.083021}. Taking $\psi_z = 0$ with different degrees of correlation between $\psi_x$ and $\psi_y$, this state leads to a novel partially polarized initial condition with non-zero $\langle S^z\rangle$. 
\end{enumerate}

The random cases (1,2) both approximate the ``unpolarised'' cosmology, with different random admixtures of longitudinal and transverse polarizations of $A_\mu$. Case (3) we simulate is a special case of the new ``mixed polarisation'', where (due to $\psi_z=0$) there is a mix of spin up and spin down randomly (like an Ising model), specified as follows:
\begin{eqnarray}
\psi_x &=& \psi_1,
\label{eq:corr_ini_x}\\
\psi_y &=& C_1 \psi_1 + C_2 \psi_2.
\label{eq:corr_ini_y}
\end{eqnarray}
with $C_{1,2}$ is the correlation parameters of polarization, with their sum is normalized ($|C_1|^2 +|C_2|^2 =1$). A fully mixed case with non-zero $\langle S^x\rangle$ and $\langle S^y\rangle$ could be obtained similarly to (3) with an independent $\psi_z\neq 0$: we don't simulate this but inferences about it follow by generalisation of our results. The mixed case (3) is a new finding with implications for experiments not discussed previously in the literature.

The fully circularly polarized case is not simulated, since it trivially has $S=1$ at all times. The fully linearly polarized case is also trivial, having just a single $\psi_i$.
 
\section{Evolution of particle number and spin density under different initial conditions}
\label{sec:Evo_number_spin_density}

In our study, the box size is chosen to be $\widetilde{L}>2\pi/\widetilde{k}_J$, where $\widetilde{k}_J=(4\widetilde{n})^{1/4}$ is the dimensionless Jeans scales, so that there is a halo formed in the box~\cite{Schive:2014dra,Eggemeier:2019jsu}. We first consider the global picture of density and spin evolution in halos. We then consider in detail the evolution of the spin vector in different regions in the dark matter halo for our different initial conditions. We use the virial theorem to find the virial radius, $r_{vir}$ of the halo, and define regions relative to this radius.


\subsection{Global Properties of Density and Spin}

Fig.\ref{fig:Slice_Rho_spin} shows slices through our simulation boxes for each initial condition, displaying the VDM density and spin. The density is shown in the left column of Fig.\ref{fig:Slice_Rho_spin}, where we observe the formation of a halo with a Proca star in its center. Due to the identical total mass conditions in CF and RF, the time for the formation of stable halos and Proca stars is roughly the same, occurring around $\widetilde{t} = 12000$. In the case of CF/$\psi_i = 0$, however, due to the smaller total mass and the correlation among the components, halos and Proca stars had already formed at $\widetilde{t} = 4500$ and $\widetilde{t} = 7500$ respectively. This is consistent with the results in Ref.~\cite{PhysRevD.108.083021}. 

\begin{figure*}[htbp]
\centering
\textbf{}\par\medskip
\begin{minipage}{0.5\textwidth}
  \centering
  \includegraphics[width=\linewidth]{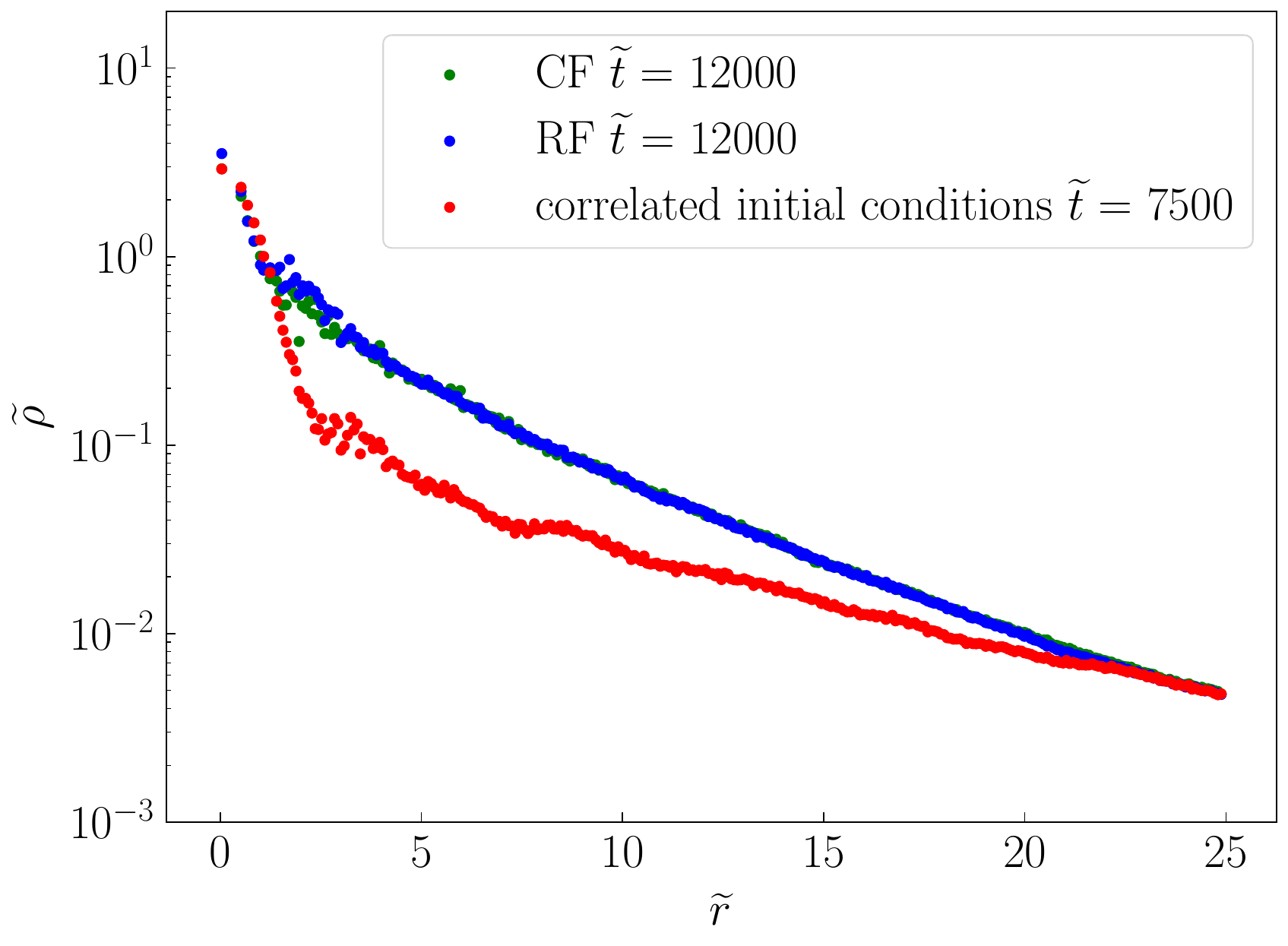}
\end{minipage}\hfill
\begin{minipage}{0.5\textwidth}
  \centering
  \includegraphics[width=\linewidth]{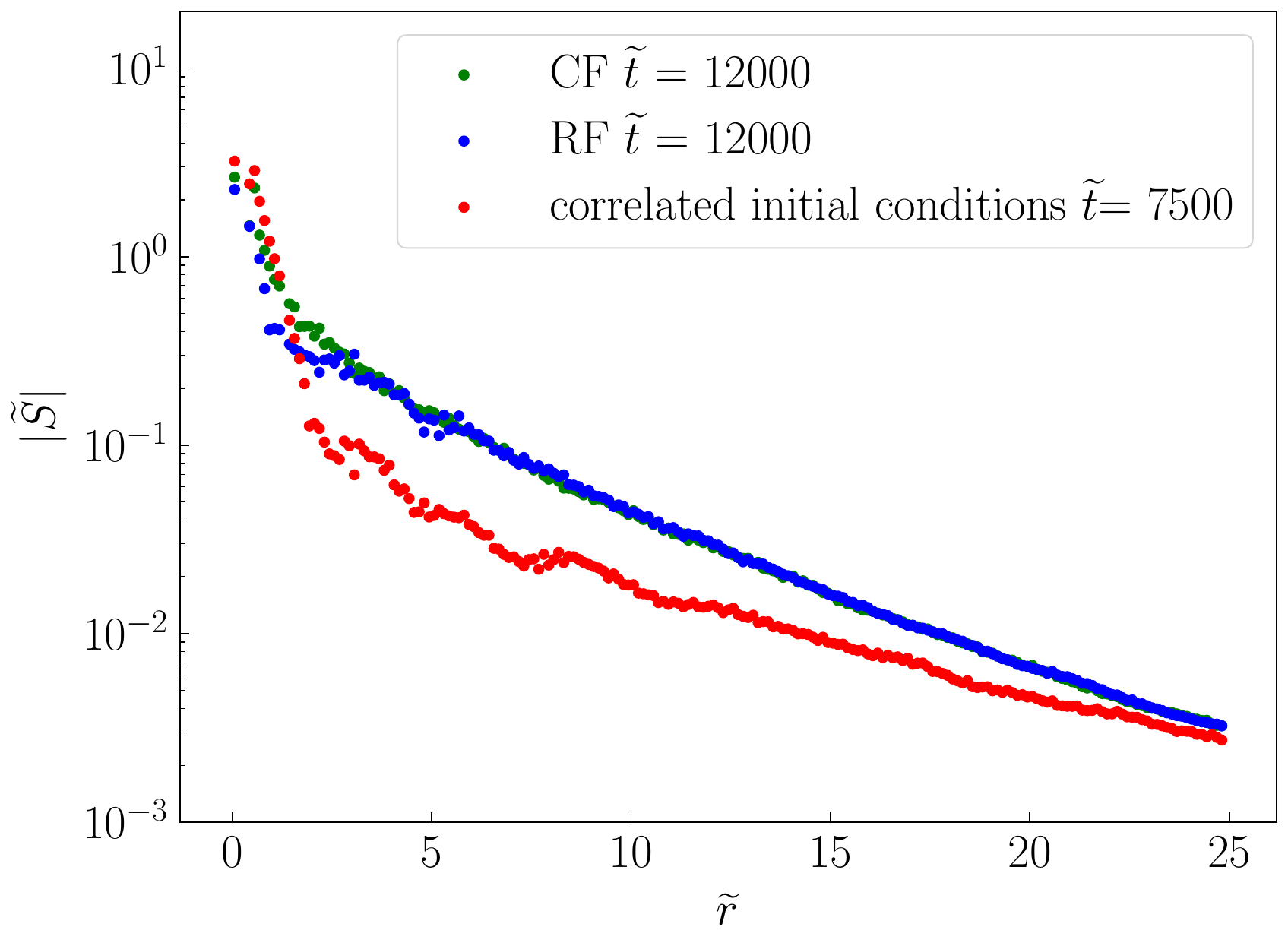}
\end{minipage}

\caption{\emph{Left:} Radial density profiles after the Proca star is formed (CF, RF at $\widetilde{t} = 12000$; correlated at $\widetilde{t} = 7500$). \emph{Right:} Spin profiles.}
\label{fig:profile_Rho_spin}
\end{figure*}

The density is shown in the right column of Fig.\ref{fig:Slice_Rho_spin}. For RF initial condition the initial spin density is everywhere zero at the initial time, see the first figure, $\widetilde{t} = 0$, in middle right column of Fig.\ref{fig:Slice_Rho_spin}. However, evolved forwards this initial condition rapidly develops random spin, since the zero state is not stable. In all cases, the spin density gathers into the halo and Proca star, a feature common for all three initial conditions, see rightmost column of Fig.\ref{fig:Slice_Rho_spin}.

The radial profiles of spin and density are shown in Fig.\ref{fig:profile_Rho_spin}, where we also observed the corresponding relationship between number density and spin density after evolution: where the number density is high, the spin density is also high, and vice versa. 

In the global view of the spin vector norm shown in Fig.\ref{fig:Slice_Rho_spin}, right panel, there does not appear to be a very large difference between the unpolarized initial conditions and the polarized ones. The difference will only become apparent when we observe the individual components of the spin, $S^i$.

\subsection{Uncorrelated initial conditions}\label{sec:No_polarization}

\begin{figure}[htbp]
\includegraphics[width=\columnwidth]{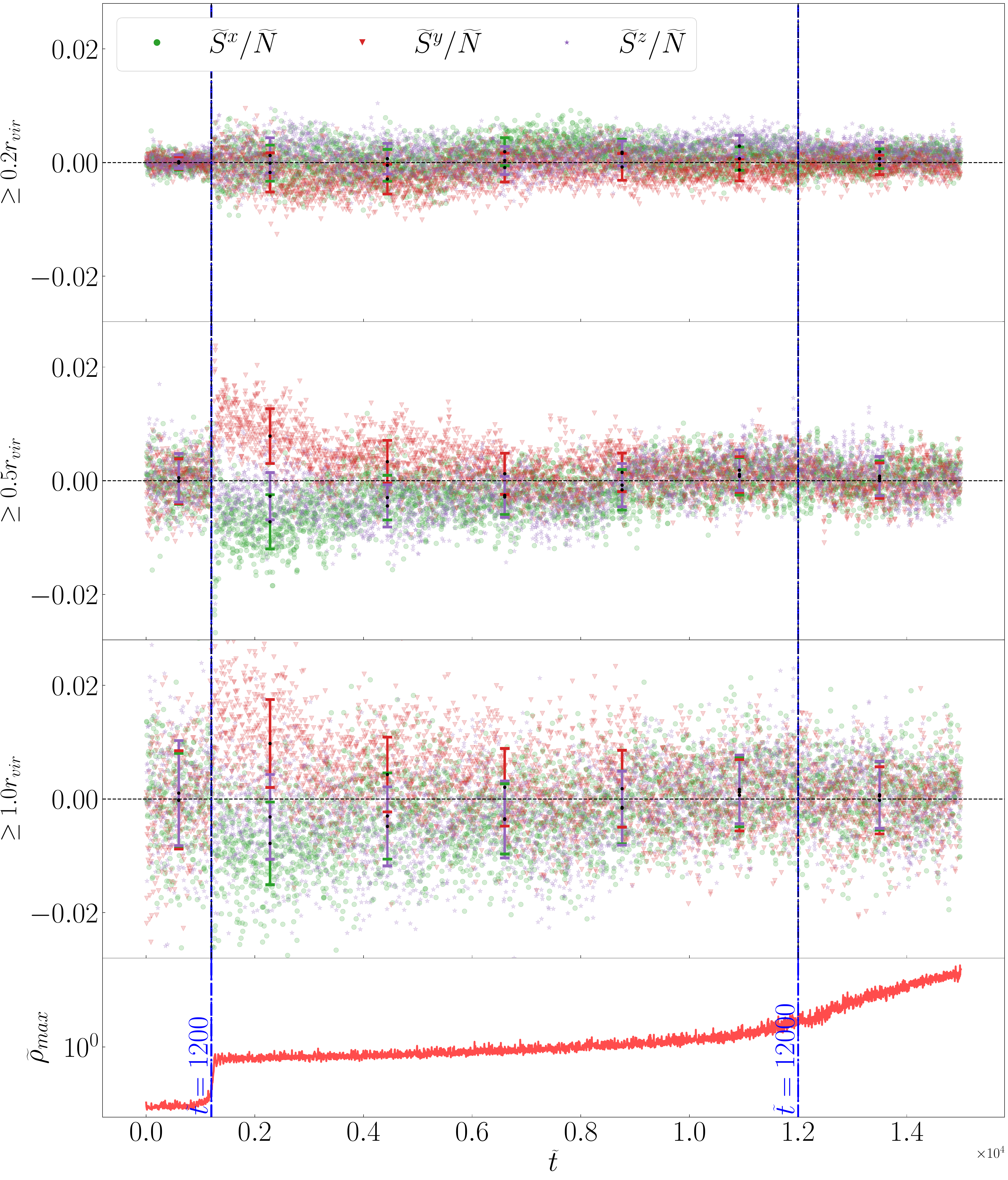}
\caption{\textbf{RF-initial condition.} The three subfigures above are the evolution of average spin density and its error bar in different regions of the halo, $\geq 0.2, 0.5, 1.0 \widetilde{r}_{vir}$. The bottom subfigure shows the growth of maximum density to indicate different phases of evolution: halo formation at $\tilde{t}=1200$ and Proca star condensation at $\tilde{t}=12000$. The blue line is $\widetilde{t} = 1200$ and $\widetilde{t} = 12000$, representing the formation of halo and proca star, respectively. Here $\widetilde{r}_{vir}\approx 25$, $\langle\widetilde{S}^i/\widetilde{N}\rangle = 0$}
\label{fig:Outer_halo_Spin_mean_evolution_real}
\end{figure}

We begin by considering the RF initial conditions, which have zero spin density everywhere at the initial time. Fig.\ref{fig:Outer_halo_Spin_mean_evolution_real} shows the evolution of the spin in each component in regions of the halo with $r > 0.2, 0.5, 1.0 \widetilde{r}_{vir}$. Throughout most of the evolution, each component of the spin individually oscillates with zero mean, with the oscillations being due to the coherence time in the halo. The only notable feature in the evolution of the halo is the short time period just after halo formation when some components of the spin take on non-zero averages. This non-linear production of short lived spin density may be of interest for VDM in unrelaxed astrophysical environments. From Fig.\ref{fig:Outer_halo_Spin_mean_evolution_real} it is also clear that the amplitude of the spin fluctuations depends on the location within the halo, with larger amplitude fluctuations outside the halo than inside.

\begin{figure}[htbp]
\includegraphics[width=\columnwidth]{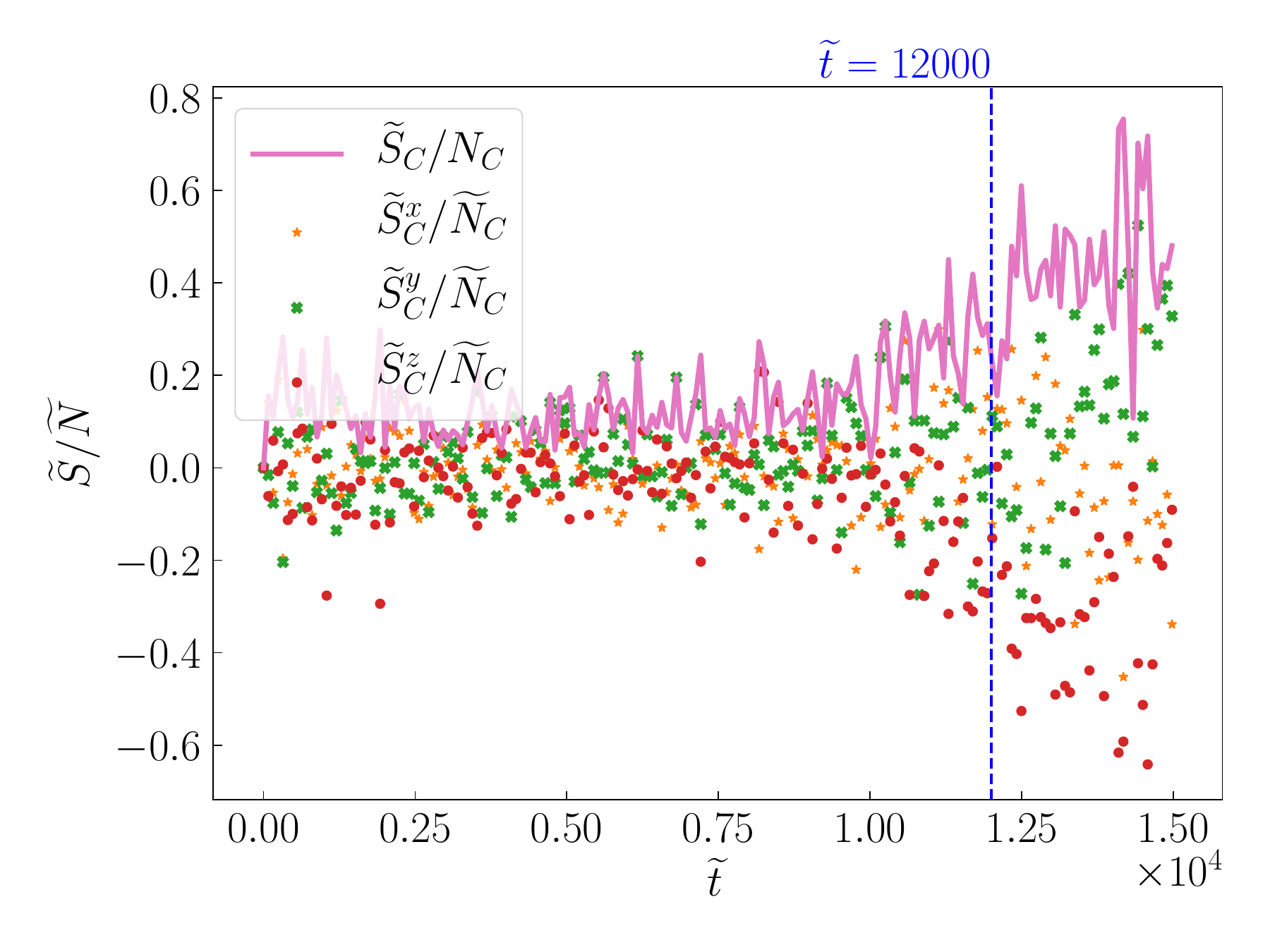}

\caption{\textbf{RF-initial condition.} Evolution of average spin density of the core with $\widetilde{N}_1=\widetilde{N}_2 =\widetilde{N}_3 \approx 630$, $\widetilde{L}=64$.The blue dashed line indicates the condensation time of the Proca star, $\widetilde{t}= 12000$.}
\label{fig:core_spin_real}
\end{figure}

Despite zero spin density on average over most of the simulation, the Proca star nonetheless develops polarization. This is demonstrated in Fig.\ref{fig:core_spin_real} which shows $\widetilde{S}^i/\widetilde{N}$ evaluated for the Proca star, we define the radius of Proca star, $r_\mathrm{core} = \max(\mathrm{density})^{-0.25} \times 1.08$~\cite{Schive:2014dra,Chen:2023bqy}. After formation, the norm of the spin develops a non-zero value, which grows and undergoes strong oscillations with time. The Proca star has $\widetilde{S}^i/\widetilde{N}<1$ and appears to approach the limiting case $\widetilde{S}^i/\widetilde{N}=1$ as time goes on. The reason is that in the case of gravity alone, different spins of solitons can overlap, and the resultant is still a soliton, depending on whether there are more directional solitons, average total spin of 0, or more spinning solitons, average total spin of 1, then the soliton formed by the superposition can take any value between 0 and 1.


\begin{figure}[htbp]
\includegraphics[width=\columnwidth]{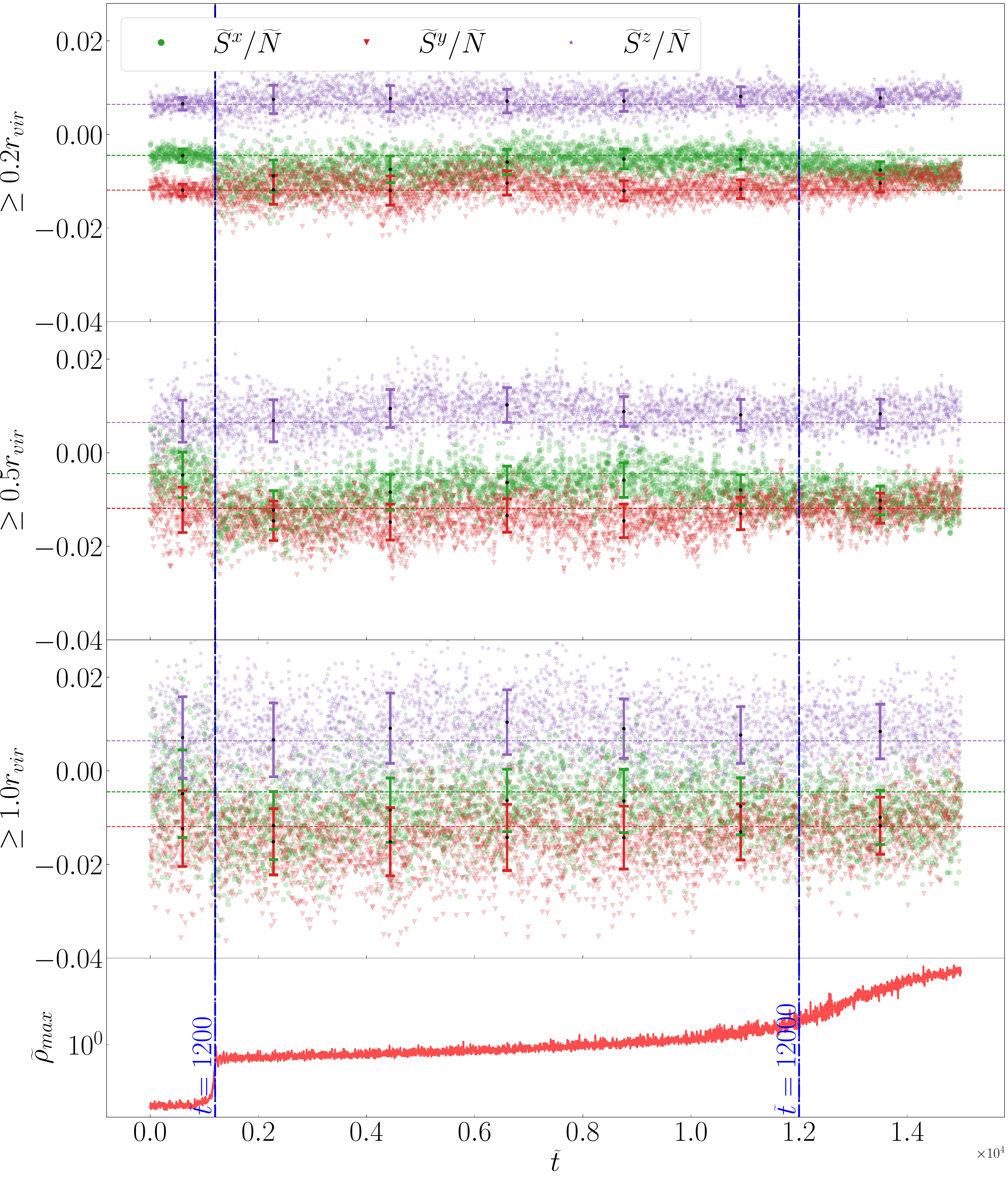}

\caption{\textbf{CF-initial condition.} The three subfigures above are the evolution of average spin density and its error bar in different area, $\geq 0.2, 0.5, 1.0 \widetilde{r}_{vir}$. The bottom subfigure shows the growth of maximum density. The blue line is $\widetilde{t} = 1200$ and $\widetilde{t} = 12000$, representing the formation of halo and proca star, respectively. Here $\widetilde{r}_{vir}\approx 25$, initial $\widetilde{S}^x/\widetilde{N} = -0.00447$, $\widetilde{S}^y/\widetilde{N} = -0.0119$, $\widetilde{S}^z/N = 0.00649$}
\label{fig:Outer_halo_Spin_mean_evolution_complex}
\end{figure}

\begin{figure}[htbp]
\includegraphics[width=\columnwidth]{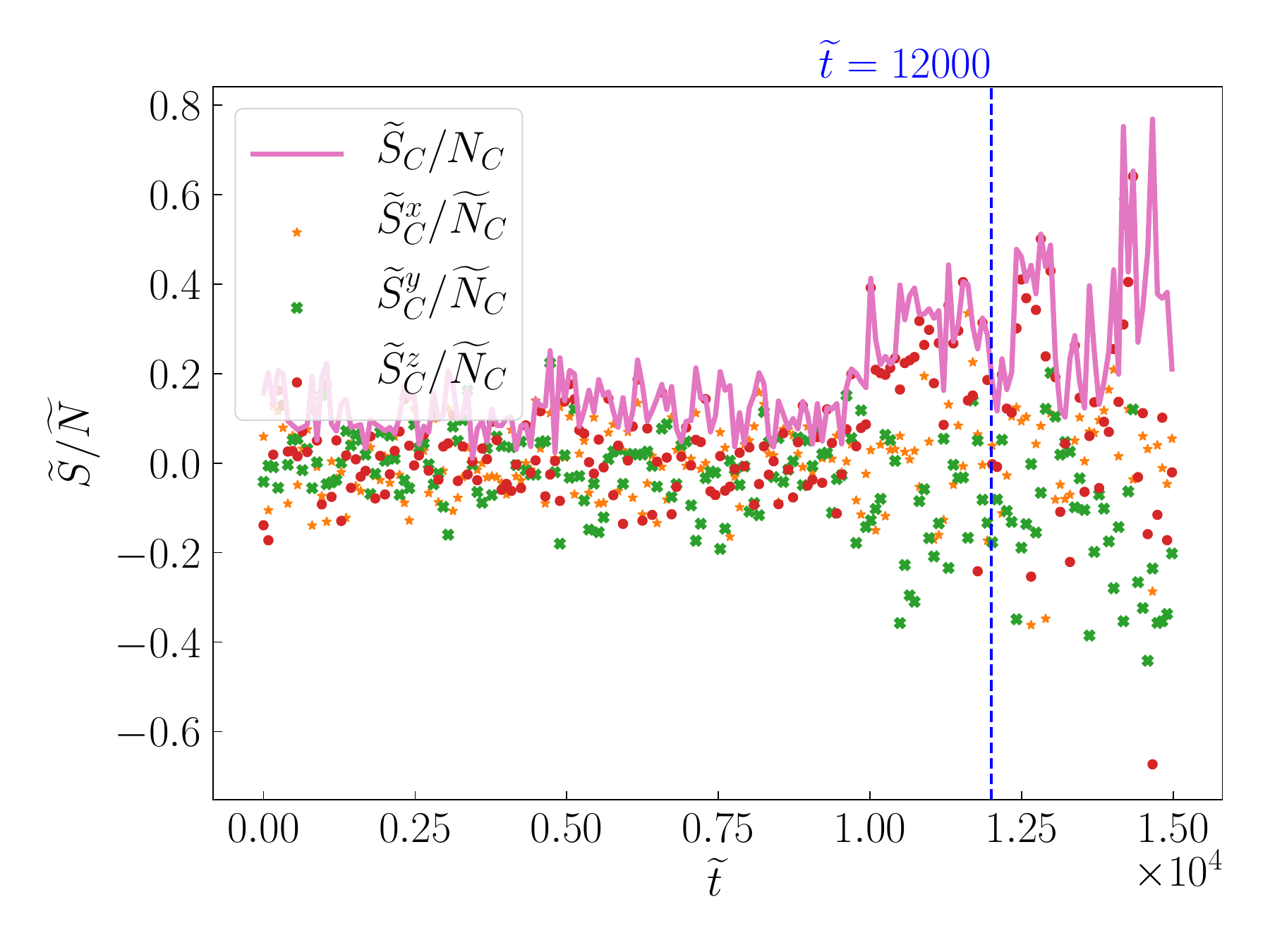}
\caption{\textbf{CF-initial condition.} Evolution of average spin density of the core with $\widetilde{N}_1=\widetilde{N}_2 =\widetilde{N}_3 \approx 630$, $\widetilde{L}=64$. The blue dashed line indicates the condensation time of the Proca star, $\widetilde{t}= 12000$.}
\label{fig:core_spin_complex}
\end{figure}

Now consider the CF initial conditions. In theory, the initial state should be completely unpolarized. However, this is only true in an infinite volume. Our initial state in fact gives rise to a significant variance in the spin components within a simulation box, and we observe the polarization at the initial state is not exactly zero but has a small value, $\widetilde{S}^x/\widetilde{N} = -0.00447, \widetilde{S}^y/\widetilde{N} = -0.0119, \widetilde{S}^z/\widetilde{N} = 0.00645$, which persists throughout the evolution. This is demonstrated in Fig.\ref{fig:Outer_halo_Spin_mean_evolution_complex}. The value of each spin component oscillates around a fixed small average everywhere in the halo. The correlation between amplitude of spin fluctuations with location in the halo is the same as in the RF case, as is the evolution of the spin of the core (Fig.\ref{fig:core_spin_complex}).

The spontaneous spin generation in the CF case is caused by the power spectrum of our initial conditions, and we have checked that the value generated is consistent with the variance on the scale of the box. This random initial spin generation can be interpreted as a generalization of the mixed polarization state, with non-zero expectation value in each component. In a cosmological setting, different Hubble volumes should give an overall spin of zero, but it could be that individual halos retain some mixed state polarisation. We will study cosmological simulations in a future work. 



\subsection{Spin polarization from correlated initial conditions}
\label{CF_psi_i_initial_condition}

\begin{figure}[htbp]
\includegraphics[width=\columnwidth]{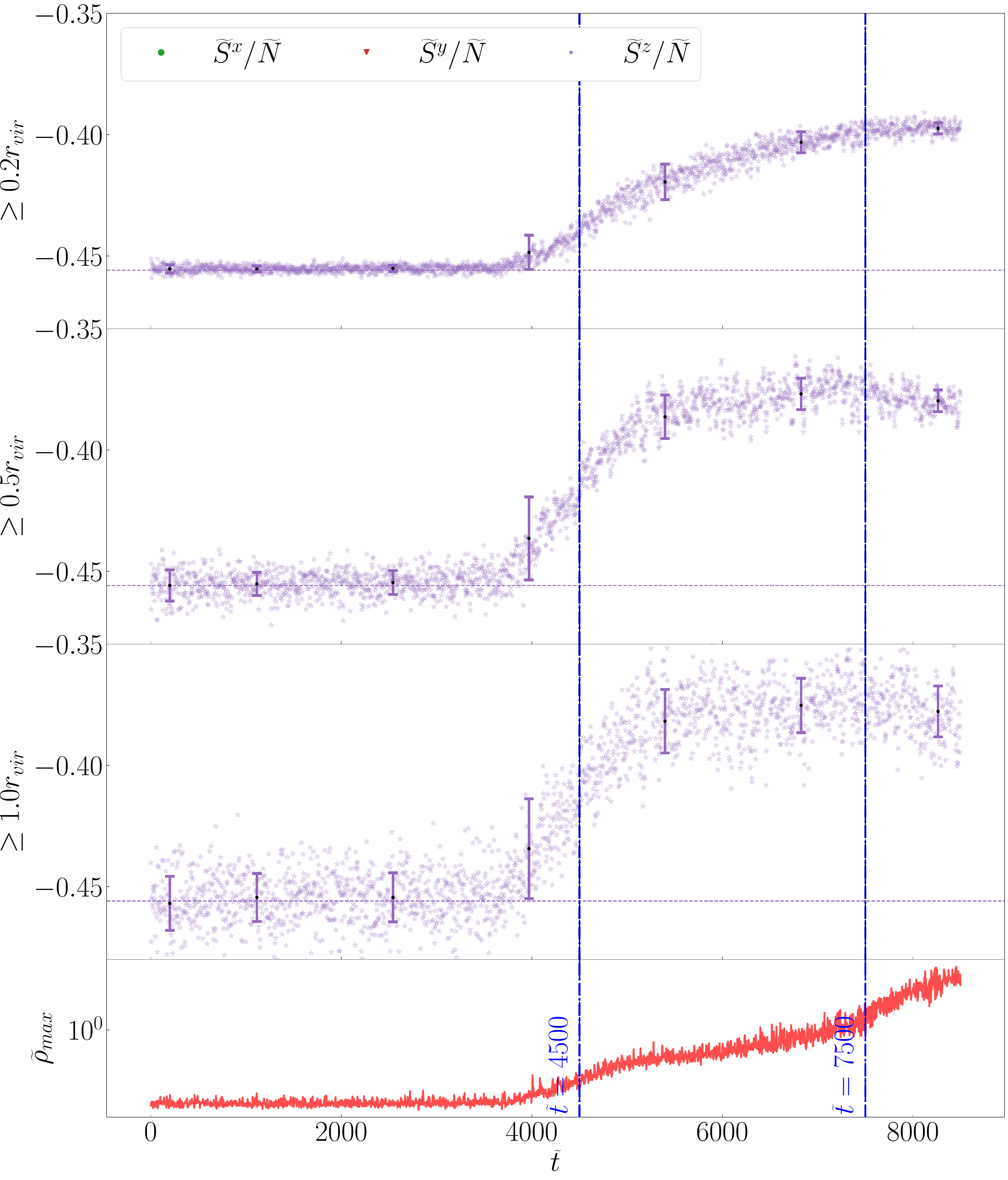}

\caption{{\bf Correlated initial conditions.} The three sub-figures above are the evolution of average spin density and its error bar in different area, $\geq 0.2, 0.5, 1.0\widetilde{r}_{vir}$. The bottom subfigure shows the growth of maximum density. The blue line is $\widetilde{t} = 4500$ and $\widetilde{t} = 7500$, representing the formation of halo and proca star, respectively.Here $\widetilde{r}_{vir}\approx 24.25$, initial $\widetilde{S}^z/\widetilde{N} = -0.456$, $\widetilde{S}^x/\widetilde{N} = \widetilde{S}^y/\widetilde{N} = 0$.}
\label{fig:Outer_halo_Spin_mean_evolution_pol}
\end{figure}

\begin{figure}[htbp]
\includegraphics[width=\columnwidth]{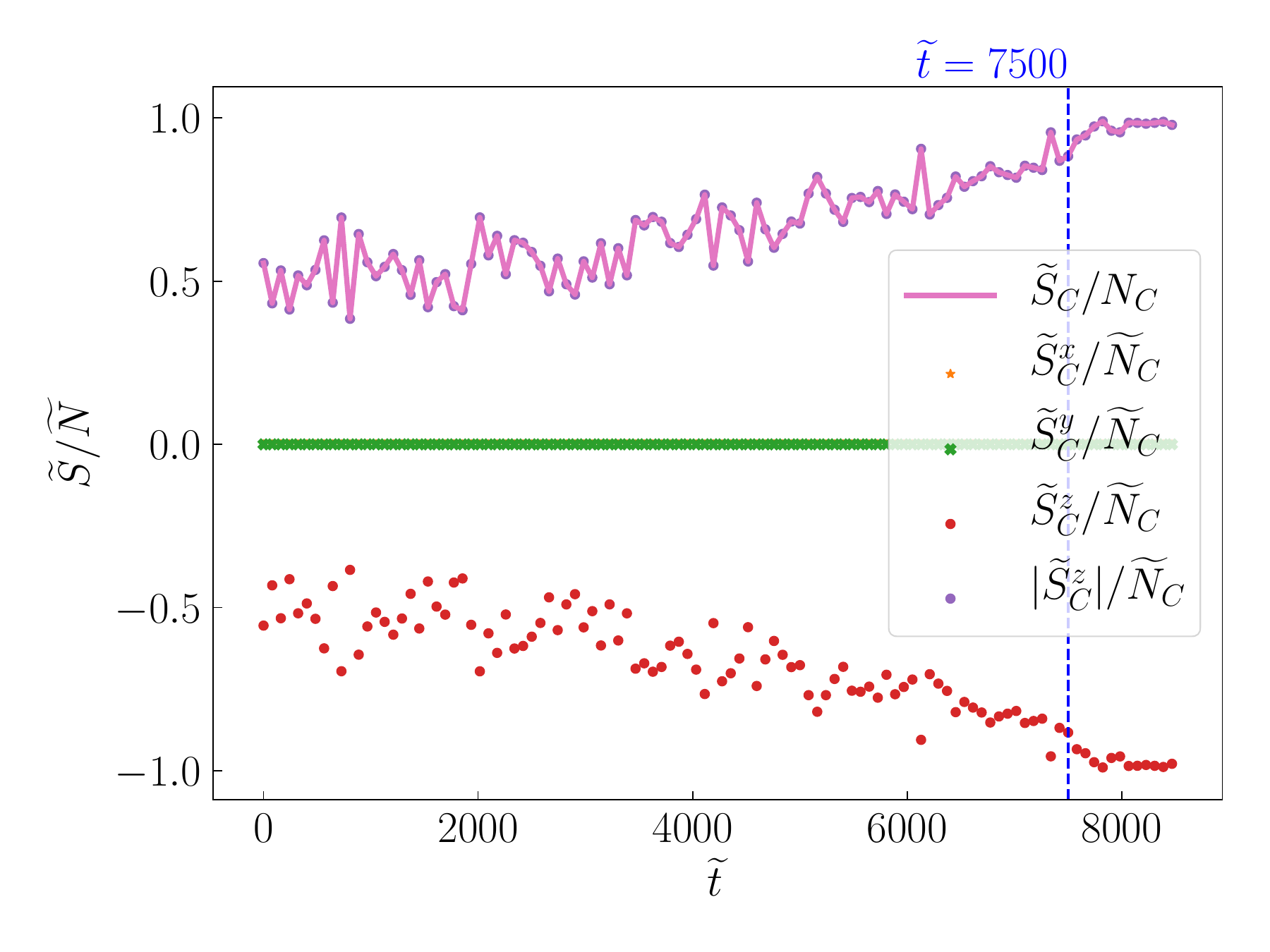}

\caption{{\bf Correlated initial conditions.} Evolution of average spin density of the core for initial conditions with $\widetilde{N}_1=\widetilde{N}_2\approx 630$, $\widetilde{N}_3 = 0$  $\widetilde{L}=64$. The blue dashed line indicates the condensation time of the Proca star, $\widetilde{t}= 7500$.}
\label{fig:core_spin_pol}
\end{figure}

Compared to the previous unpolarized state where each component $\psi_i$ of the system was independent in the initial condition, we now set $\psi_z=0$ and correlate $\psi_{x,y}$ using Eqs.~(\ref{eq:corr_ini_x}, \ref{eq:corr_ini_y}). This gives a partially circularly polarized state initially. By employing Eq.~\ref{eq:conser_S}, it is straightforward to deduce that under these conditions only the $S^z$ component possesses spin. With a high degree of correlation, we enforce a large and resolvable (larger than numerical error) value to $S^z$. Symmetry dictates that the other components of the spin remain exactly zero everywhere, which the system respects. The spin in the halo is shown in Fig.\ref{fig:Outer_halo_Spin_mean_evolution_pol}, while the spin of the core is shown in Fig.\ref{fig:core_spin_pol}.

The polarized initial condition we have chosen does not lead to an absolutely fixed spin, but the spin evolves when the halo is formed at $\tilde{t}=4500$, and undergoes fluctuations within the halo. As with the other cases, we observe that the spin fluctuation amplitude is larger in the outer ranges of the halo. The core is formed at $\tilde{t}=7500$ and is born with near maximal spin. 

The correlated state that we have simulated here is a new state of VDM, with non-maximal spin polarzation and superimposed fluctuations. There are obvious generalizations of it, with the correlation matrix chosen to give large averages for each $\widetilde{S}^i$, along with fluctuations. We now discuss the physical interpretation and experimental implications of mixed state VDM polarization.

\section{Spin Mixed State of VDM}\label{sec:spin_density_and_mixing state}

To understand the result on the average spin density ($\widetilde{S}/\widetilde{N}$) in different regions of the halo under the initial conditions discussed in Sec.~\ref{sec:Evo_number_spin_density}, we propose the ``mixed state'' between random and aligned distributions of the spin vector of VDM. The approach aims to explain the collective behavior of the VDM spin in different regions of the halo. By understanding this collective behavior of the VDM spin vector, we can bridge to the implications for direct searches using haloscopes discussed in the next section. 

\begin{figure}
    \centering
    \includegraphics[width=1\linewidth]{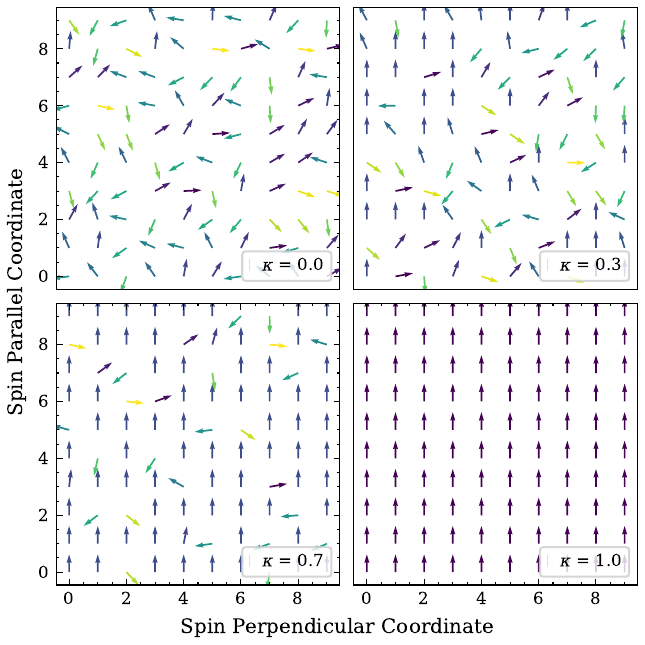}
    \caption{Visualization of the mixed state of the VDM with different configurations of $\kappa$. The amplitude of each spin vector is one, and the varying colors indicate different pointing angles of the spin vector. For visualization purposes, we simplify the distribution of the spin vectors from 3D to 2D and fix the orientation of the spin-aligned component to the y-axis.}
    \label{fig:mix_state}
\end{figure}

The mixed state of VDM is composed of a random spin component and an aligned spin component. Within the interested volume $V$, there are $N_\mathrm{VDM}$ VDM dark matter particles. We define the factor $\kappa$ that characterizes the composition of the aligned component given by:
    \begin{equation}
        N_\mathrm{VDM} = \underbrace{\kappa N_\mathrm{VDM}}_\text{aligned spin} + \underbrace{(1-\kappa) N_\mathrm{VDM}}_\text{random spin} \label{eqn:mix-state}
    \end{equation}
    At first glance, the mixed state takes a formulation similar to the Ising model used to describe ferromagnetism in materials, albeit without the spin interaction term. Fig.\ref{fig:mix_state} presents how the mixing degree of VDM depends on the various values of $\kappa$. 
    \begin{enumerate}[i]
        \item  $\kappa = 0$: The VDM population is composed of random spin particles, thus the average spin component ($\widetilde{S}^i/\widetilde{N} = 0$) is trivially zero. Such a condition exists in the case of uncorrelated initial conditions, i.e., CF and RF cases, in the outer halo region (Figures~\ref{fig:Outer_halo_Spin_mean_evolution_real} and~\ref{fig:Outer_halo_Spin_mean_evolution_complex}), and at the very early stages of the core's evolution timeline (Figures~\ref{fig:core_spin_real} and~\ref{fig:core_spin_complex}). 

        \item $0 < \kappa < 1$: The VDM is composed of both random spin and spin-aligned particles. Thus, the average spin density of at least one spatial component ($\widetilde{S}^i$) is non-zero, and the \textit{average spin amplitude is also equal to the factor $\kappa$ itself}. The increase of the mixing factor $\kappa$ during the evolution of the core (Figures~\ref{fig:core_spin_real}, \ref{fig:core_spin_complex}, and \ref{fig:core_spin_pol}) shows that during the condensation of the halo until the formation of the Proca star, the VDM population at the core region gradually becomes coherent in spin.

        \item $\kappa = 1$: The VDM population is composed of spin-aligned VDM particles, meaning at least one average spin component is non-zero, and the average spin amplitude is equal to one ($\widetilde{S}/\widetilde{N} = 1$), as depicted in the bottom-right sub-figure of Fig.\ref{fig:mix_state}. The environment where $\kappa = 1$ can exist in the core region where the spin density asymptotically increase to one as the simulation time reaches very large values (see Fig.\ref{fig:core_spin_pol}). VDM could also have an initial condition where the correlation is highest, resulting in an initial spin density equal to one, e.g., $C_1 = 1/\sqrt{2}, C_2 = i/\sqrt{2}$ \cite{PhysRevD.108.083021}. 
    \end{enumerate}

    Throughout the results presented in Sec.~\ref{sec:Evo_number_spin_density}, it can be observed that there is a minor degree of fluctuation in the average spin components, approximately 1-2\%, in the far outer region of the halo ($\widetilde{r} \sim \widetilde{r}_\mathrm{vir}$). It can be seen that the fluctuation's deviation decreases when the region of interest is extended closer to the center of core (Figures~\ref{fig:Outer_halo_Spin_mean_evolution_real} and~\ref{fig:Outer_halo_Spin_mean_evolution_complex}). Under the initial condition of the RF and CF, there is at least one component being non-zero. This indicates the direction of the spin-aligned population varies over time, given by the virial time scale $\tau_\mathrm{vir} \approx \lambda_\mathrm{de Broglie}/v_\mathrm{vir}$. Meanwhile, for the correlated initial condition, fluctuations from random spin component exists but are dominated by the large population of spin-aligned VDM. 

    Interestingly, one also sees that the average spin amplitude increases over time at the center of the halo or Proca star. Thus, the ratio of the spin-aligned population increases during the lifetime of Proca star formation. However, in the correlated initial condition, the average spin density ($\widetilde{S}/\widetilde{N} = \widetilde{S^z}/\widetilde{N}$) in the outer halo region decreased in sync with the increase of spin density in the halo's center, resulting from the conservation of spin momentum $\int_V \widetilde{S^z}dV = \mathrm{const}$. Thus, we also expect that if the system evolves long enough ($t \rightarrow \infty$), the VDM population of the Proca star will be spin-aligned dominated, and the ratio of this population in the halo would become the lowest. 
    
    In summary, Tab.~\ref{tab:mix_state} summaries the VDM dynamics under various initial conditions in two regions of interest. The terms ``Aligned'' and ``Random'' describe the spatial mixing state, dominated by either aligned or random spin populations of VDM. Meanwhile, ``Varying'' and ``Stationary'' indicate the time evolution of the spin vector associated with the spin-aligned component.
    
\begin{table*}[]
\begin{tabular}{lc||cc}
 & \multicolumn{1}{l||}{} & \multicolumn{2}{c}{\textit{\textbf{Region of Interest}}}                           \\
 & \multicolumn{1}{l||}{} & \multicolumn{1}{c|}{Halo Center/Proca Star}                    & Outer Halo               \\ \hline \hline
\multirow{3}{*}{\textit{\textbf{Initial Condition}}} & CF           & \multicolumn{1}{c|}{Aligned - Varying}  & Random  - Varying             \\ \cline{2-4} 
 & RF                    & \multicolumn{1}{c|}{Aligned - Varying} & Random - Varying  \\ \cline{2-4} 
                                                    & CF/$\psi_z=0$ & \multicolumn{1}{c|}{Aligned - Stationary } & Aligned  - Stationary 
\end{tabular}
\caption{Summary of VDM's dynamics to the initial conditions and regions of interest. Each cell shows the dominant type of spatial distribution, characterized by either fixed or random polarization spin configurations, as well as their directional dynamic (varying direction or stationary direction).}\label{tab:mix_state}
\end{table*}

\section{Implication to Direct Searches}\label{sec:direct_searches}

Experiments that directly search for VDM, called \textit{haloscopes}, utilize the conversion of VDM particles to photons within the geometry of the experiment, utilizing the kinetic mixing coupling between the detectable electric field and the dark (or ``hidden'') electric field of the VDM. Thus, in this section, we focus on the behavior of the detectable signal power of VDM within the haloscope. The interpretation is based on the dynamics of VDM in the outer halo region (Sec.~\ref{sec:Evo_number_spin_density}, in a location that is geometrically comparable to the location of the Solar System in the Milky Way. Starting from the properties of the spin of VDM particles, one can obtain the behaviour of the dark electric field, $\bm{E}\approx \partial_t\bm{A}$, of VDM given by the following equation \cite{Amaral:2024tjg}:
\begin{equation}
    \bm{S}  = \bm{A}\times \dot{\bm{A}}\, .
\end{equation}
Thus, non-zero spin defines \emph{circular} polarization of the dark electric field. The rotation period of the polarization is given by the Compton period of the VDM ($T_\mathrm{C} = 1/m_\mathrm{VDM}$). This time scale is also the period of the detectable EM signal in the haloscope experiment. Circular polarization (as emphasised by e.g. Refs.~\cite{Amin:2022pzv,Amaral:2024tjg}) presents a stark contrast with the behavior of $\bm{A}$ widely assumed by the VDM haloscope search community, which is \textit{linear polarization} i.e., $\bm{A}$ of a VDM particle oscillates in one direction, so $\bm{S} = 0$. The community generally assumes linear polarization for $\bm{A}$, where the \textit{random polarization} case is equivalent to the random distribution of the direction of the dark electric field, while \textit{fixed  polarization} over long time scales (much larger than sidereal period) corresponds to the aligned direction of the dark electric field~\cite{Arias:2012az}. Ref.~\cite{Caputo:2021eaa} discusses the sidereal modulation of the VDM signal in the latter case of fixed polarization. This linear polarization condition can only exist under very specific conditions of the VDM, which result in $\widetilde{S}=0$ and differs from our findings of non-zero $\bm{S}$ for each VDM particle. 

Nevertheless, the results from the linear polarization of the dark electric field can be improved to interpret the behavior of the signal from haloscopes in the scenario of circular polarization. The signal power of a VDM haloscope is given by 
    \begin{equation}
        P_\mathrm{VDM} = \chi^2 \rho_\mathrm{DM} \mathcal{G} \alpha^2 \, ,
        \label{eqn:signal_power}
    \end{equation}
    with $\chi$ (or $\epsilon$) as the kinetic mixing factor between VDM and photons, $\rho_\mathrm{DM}$ as the local dark matter density that relate to the dark electric field by the relation~\cite{Arias:2012az,Horns:2012jf}
    \begin{equation}
        \rho_\mathrm{DM} = \frac{m^2_\mathrm{VDM}}{2} |\bm{A}|^2
    \end{equation}
and $\mathcal{G}$ as the factor that depends on the geometrical design and coupling efficiency of the signal to the detector, e.g., for resonant cavity experiments, $\mathcal{G}$ depends on the parameters of the cavity such as quality factor ($Q$), volume ($V$), and the form factor of the detection's mode $(\mathcal{C}_{nml})$ \cite{Kimball_book_2023}. Meanwhile, $\alpha^2$ is the fraction of the dark electric field vector of the VDM that is \textit{converted} and \textit{detected} by the haloscope. For example, in the concept of a dish antenna \cite{Horns:2012jf}, the VDM-photon conversion on the conducting surface depends on the normal vector of the surface ($\hat{\bm{G}}$) and the polarization of the VDM, given by the term $1 - (\hat{\bm{A}}\cdot \hat{\bm{G}})^2$. The choice of detector determines the fraction of detectable signal: a single polarization antenna coupled with a  linear detector can only detect one component of the electric field parallel to its polarization, while a square-law detector\footnote{power detector or incoherent detector} can register the full signal power of the EM signal. 


    \subsection*{Modulation timescale: Compton and Sidereal}
    
Concerning the impact of the mixed state on the distribution of the dark electric field, random spin VDM will result in a random distribution of the dark electric field. If random spin VDM dominates, $\kappa = 0$, and the coherence time scale ($\tau_\mathrm{c}$) is much shorter than the observation time ($\tau_\mathrm{obs}$), one could expect the distribution of the dark electric field to be similar to the case of linear random polarization~\cite{Arias:2012az}. Ref.~\cite{Amaral:2024tjg} discusses the case of very low mass VDM (e.g., $m_\mathrm{VDM} < 1 \text{ Hz}$), where the coherence time scale is comparable to the observation time scale ($\tau_\mathrm{c} \approx \mathcal{O}(\tau_\mathrm{obs})$). In this scenario, the signal detected using an accelerometer-based network will show a strong dependency on the polarization rotation enveloped within the sidereal time scale.

In an aligned spin VDM population, either from the extreme aligned-spin condition or mixed state, the spin vector of this population is fixed over the universal time scale shown in Fig.~\ref{fig:Outer_halo_Spin_mean_evolution_pol}. This leads to the result that the rotational plane of the dark electric field vector is perpendicular to the spin vector $\bm{S}$ and fixed over the same time scale of $\bm{S}$. The dark electric field will coherently rotate and collectively point to the same direction. In comparison with the study carried out by Ref.~\cite{Caputo:2021eaa} for linear polarization, the output signal of the haloscope will show modulation characterized by the Compton period due to the circular polarization, and further rotation of the Earth will cause sidereal modulation of the signal. 

\begin{figure}
        \centering
        \includegraphics[width=1\linewidth]{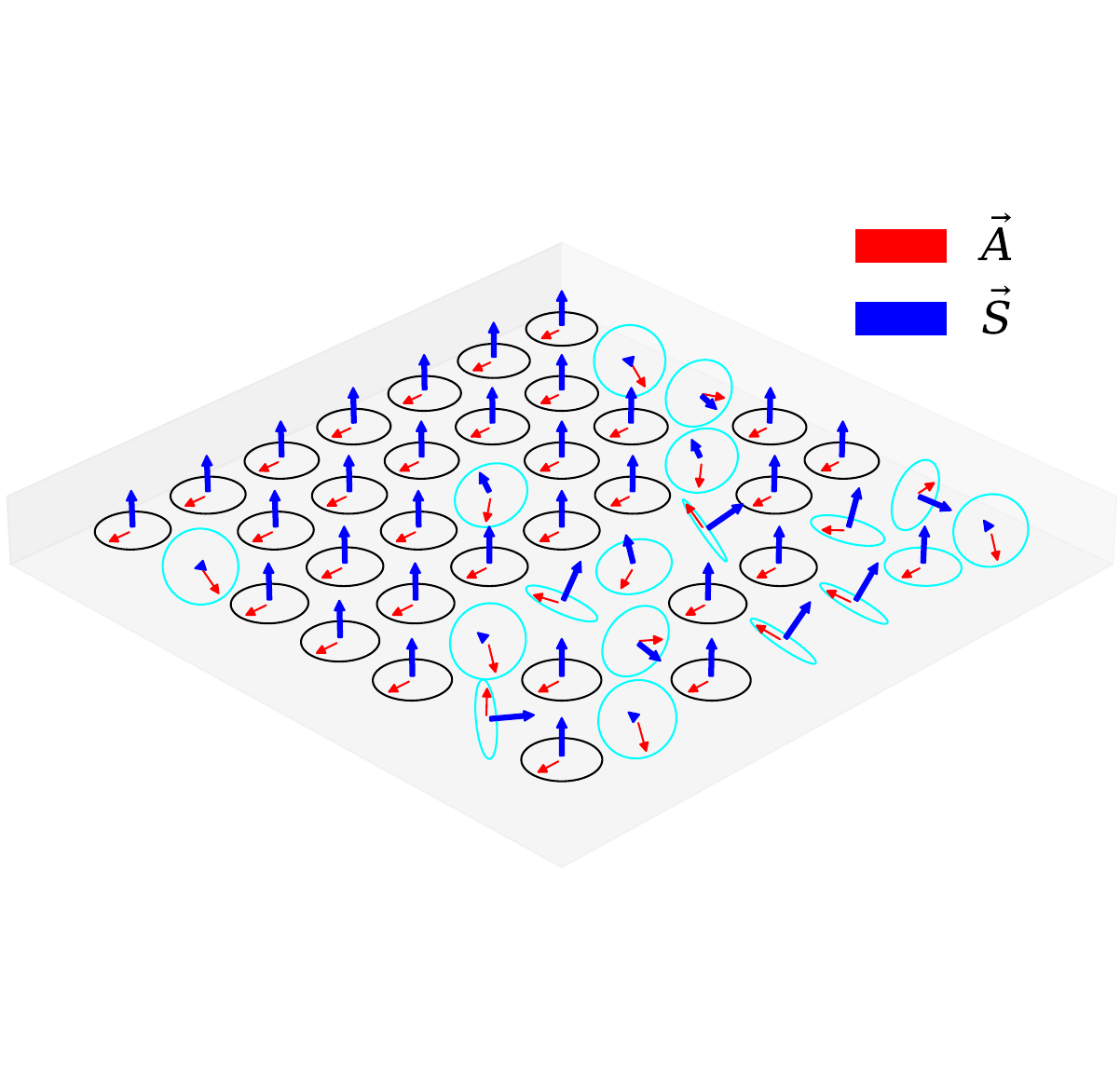}
        \caption{Distribution of the dark electric field $\bm{A}$ under the spin's mixed state. The cyan circles denote the random spin population of VDM, which results in randomly distributed $\bm{A}$. The black circles denote the aligned spin population, resulting in $\bm{A}$ collectively pointing in the same direction and coherently rotating. The $\bm{A}$ of each VDM particle has a frequency of rotation determined by its mass. }
         \label{fig:modu_brass}
    \end{figure}

Fig.~\ref{fig:modu_brass} visualizes the spin mixed state of VDM with $\kappa = 0.5$. The random spin component yields a randomly distributed dark electric field, while the aligned spin component gives rise to an aligned and coherent rotation of the dark electric field. 
To characterize the power of the haloscope in the presence of the mixed state and circular polarization, we compute the $\alpha^2$ term in Eq.~\ref{eqn:signal_power}:
    \begin{equation} 
        \alpha^2 = \begin{cases} \kappa (\hat{\bm{A}}\cdot \hat{\bm{G}})^2 + \frac{(1-\kappa)}{3}  & \mbox{axial} \\ 1- \kappa (\hat{\bm{A}}\cdot \hat{\bm{G}})^2 - \frac{(1-\kappa)}{3}  & \mbox{planar} \end{cases} \label{eqn:alpha2}
    \end{equation}
    with $\hat{\bm{G}}(T_\mathrm{S})$ having the period of sidereal time for the general case. The $\hat{\bm{G}}$ vector is the unit vector of the haloscope's sensitive axis (``axial") or the normal vector of the conversion surface (``planar")~\cite{Caputo:2021eaa}. The term $(\hat{\bm{A}}\cdot \hat{\bm{G}})^2$ exhibits modulation in both Compton and sidereal time scale caused by the polarization rotation of $\bm{A}$ and Earth's rotation effect on $\bm{G}$. During the polarization rotation's period, $\hat{\bm{A}}$ could turn perpendicular to $\hat{\bm{G}}$ (twice) thus the dot product $(\hat{\bm{A}}\cdot \hat{\bm{G}})^2$ is reduced to zero. On the other hand, $\hat{\bm{A}}$ could also aligned and be co-planar with $\hat{\bm{G}}$ and $\hat{\bm{S}}$. This configuration yields the highest possible value of $(\hat{\bm{A}}\cdot \hat{\bm{G}})^2$. The average of this term over the Compton time scale is simply
    \begin{eqnarray}
        \langle (\hat{\bm{A}}\cdot \hat{\bm{G}})^2 \rangle &=& \mathrm{max}\frac{(\hat{\bm{A}}\cdot \hat{\bm{G}})^2}{2} \nonumber \\ &=& \frac{1-(\hat{\bm{S}}\cdot \hat{\bm{G}})^2}{2} \label{eqn:dotproduc2}
    \end{eqnarray}
    
    As we will discuss in the case study below, the averaging is only applicable for specific types of haloscope where the Compton time scale is much smaller than the time scale of the data taking period. The combination of expression~\ref{eqn:alpha2} and~\ref{eqn:dotproduc2} can be used to discuss the modulation of signal by the sidereal time scale in the mixed configuration between random spin and aligned spin of VDM. 
    
    \subsection*{Case Study with BRASS-p}\label{sec:case_study}
    To demonstrate with a case study, we choose the BRASS-p experiment \cite{Nguyen:2023uis} due to its world-leading sensitivity for VDM within the mass range of (41.35 - 74.44) $\mu$eV. BRASS-p is equipped with a state-of-the-art radio telescope's full-fledged analog and digital backend, allowing it to measure the electric field using its double linear polarization receiver. As a result, the signal recovered is not only equal to the power detected by the square-law detector, but the GPU-powered digital backend can also dissect the incoming signal to obtain the four Stokes parameters of the polarization signal\cite{Robishaw:2018ylp}.
     In this case study, the chosen spin-vector of the aligned spin population is $\hat{\bm{S}} = (1/4,\sqrt{3}/4,\sqrt{3}/2)^T$, $\hat{\bm{A}} (t=0)$ can be chosen arbitrarily and $\hat{\bm{S}} \perp \hat{\bm{A}}$. Time evolution of $\hat{\bm{A}}$ in Compton period can be obtain by the Rodrigues' rotation formula around the vector $\hat{\bm{S}}$
     \begin{equation}
         \hat{\bm{A}}(t) = R_{\hat{\bm{S}}} \left( \frac{2\pi t}{T_\mathrm{C}} \right) \hat{\bm{A}} (t=0)
     \end{equation}
    
    Fig.~\ref{fig:modu_brass} shows the sidereal modulation of $(\hat{\bm{A}}\cdot \hat{\bm{G}})^2$ for the aligned spin VDM in the sidereal and Compton time scales. The polarization rotation yields a conversion rate that oscillates at the Compton time scale between zero (when $\hat{\bm{A}} \perp \hat{\bm{G}}$) and the upper value given by the configuration of $\hat{\bm{G}}$ in the sidereal time scale. Haloscope experiments typically involve long data-taking campaigns composed of many short periods ($\gg T_\mathrm{C}$), which are combined to produce a single spectrum. As a result, signal modulation on the Compton time scale is averaged out in the average spectrum. The averaging of the $(\hat{\bm{A}}\cdot \hat{\bm{G}})^2$ within the Compton period is a suitable way to present the signal per instance of data-taking, and thus the signal in the average spectrum is only subjected to the sidereal modulation. Because of this averaging in the Compton time scale, the modulation span appears to be milder than the linear polarization discussed in~\cite{Caputo:2021eaa}. As a side note, in haloscope experiments sensitive to very low-mass VDM where the coherence time $T_\mathrm{C}$ is comparable to or greater than the observation time $T_\mathrm{obs}$ or the sidereal time, the signal power depends on both $T_\mathrm{C}$ and $T_\mathrm{obs}$.

    \begin{figure*}
        \centering
        \includegraphics[width=1\linewidth]{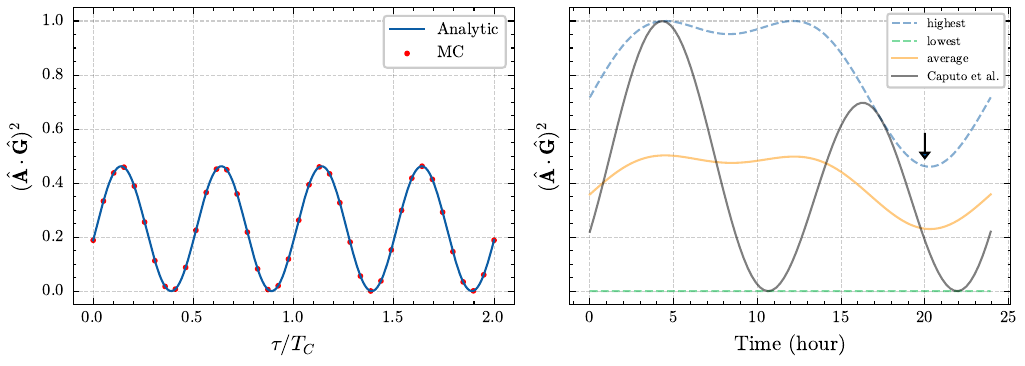}
        \caption{Modulation of the $(\hat{\bm{A}} \cdot \hat{\bm{G}})^2$ in the Compton and Sidereal time scales. \textbf{Left}: The modulation over two Compton periods observed at the 20th hour of sidereal time (indicated in the right figure). The results are obtained from both analytical calculation and Monte Carlo simulation. \textbf{Right}: The modulation over the sidereal time scale, indicating the highest, lowest, and average values from the Compton modulation for different configurations of $\hat{\bm{G}}$ as it changes over sidereal time.} 
    \end{figure*} 

\begin{figure}
        \centering
        \includegraphics[width=1\linewidth]{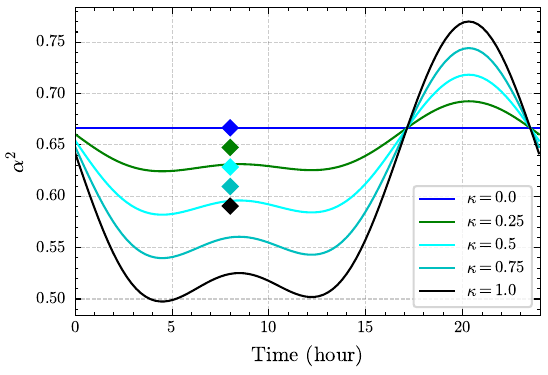}
        \caption{The sidereal modulation of $\alpha^2$ with various VDM mixed states defined by $\kappa$ in the BRASS-p experiment. The diamond markers show the average power during a sidereal period of the modulation.}
         \label{fig:sidereal_modu}
    \end{figure}

    Since BRASS-p is a planar type haloscope, by combining the equations~\ref{eqn:alpha2} and~\ref{eqn:dotproduc2}, one could obtain the sidereal modulation of the signal with different mixing ratio $\kappa$ (see Fig.\ref{fig:sidereal_modu}). One can see that the modulation reduces to the value of $2/3$ for fully random spin VDM, while it becomes most significant when the VDM population is dominated by spin-aligned components.

    In the scenario of VDM signal detection, i.e., a high signal-to-noise ratio (SNR) channel in the grand spectrum and a well-resolved Maxwellian signal profile in the vertical combined spectrum, a longer data-taking campaign carried out over an extended period could reveal the shape of the SNR's modulation, knowing that $\mathrm{SNR} \propto \alpha^2$. The shape of the modulation can be used to infer the spin-vector direction. Even in the case of the CF-initial condition where the population ratio of the fixed polarization component is only around 1\% (Fig.\ref{fig:Outer_halo_Spin_mean_evolution_complex}), a strong SNR signal (i.e., $\mathrm{SNR} = 100$) could be sufficient to reveal the modulation. After obtaining the information on the spin-vector direction, Monte Carlo simulations with different proportions of mixed states should follow to estimate the SNR modulation's deviation from the mean SNR collected during the entire campaign.
    Ultimately, exclusion limits or discovery thresholds for VDM haloscope searches should be refined to account for the full range of possible mixed states $\kappa$ of VDM. This spans from $\alpha^2 = 1/3$ or $2/3$ for fully random-spin VDM to the sidereal-modulated value of $\alpha^2$ for fully spin-aligned VDM, characterized by the vector $\hat{\bm{S}}$. Additionally, careful consideration must be given to the characteristic vector of the haloscope, $\hat{\bm{G}}$, which depends on the detector type, orientation, and location of the haloscope.

\section{Discussion and Conclusions}\label{sec:conclusion}

We have studied the spin polarization distribution and evolution in dark matter halos in fully non-linear, nonrelativistic, numerical simulations of vector dark matter (VDM). Our work builds on recent studies such as Refs.~\cite{PhysRevD.108.083021,Jain:2023ojg,Amaral:2024tjg}. We simulated different stochastic initial conditions, and studied the spin polarization in the regime of time scales much longer than the coherence time. We found that non-linear gravitational evolution does not change the spin distribution in the VDM halo on average compared to the initial conditions, due to conservation laws being respected. The spin in local regions, however, undergoes fluctuations, with a stronger amplitude of fluctuation in the outer regions of the halo compared to near the core. The core has large spin in all initial conditions. 

We checked that the halos in our simulations have angular momentum, $L$, but that there is no spin-orbit coupling and thus $L$ and $S$ of the core are not aligned in general. Spin-orbit coupling in VDM is generated by relativistic corrections~\cite{Cashen:2016neh}, which should be small inside DM halos. Spin-orbit coupling at the level of the Schr\"{o}dinger-Poisson equations can be modelled as in Ref.~\cite{Jain:2023qty}.

Our random initial conditions in two cases mimic the production mechanisms for VDM that produce either transverse or longitudinal polarizations, or admixtures thereof~\cite{Graham:2015rva,Agrawal:2018vin,Dror:2018pdh,Long:2019lwl}. One case we studied, the real field, RF, has perfect $\langle S\rangle = 0$ by symmetry. With a random complex field, CF, a small value of $\langle S/N\rangle $ is generated at random in the simulation box from the initial power spectrum, and would average out in an ensemble of simulations with different random seeds (or in a cosmological setting of disconnected causal regions). The power spectrum gives rise to mixed state polarizations of small spin.

We also proposed and simulated a novel mixed state of VDM, which mimics VDM produced by isotropy-violating processes~\cite{Arias:2012az,Alonso-Alvarez:2019ixv}, but with a preferred axis of circular, rather than linear, polarization. This mixed case has random fluctuations on top of a non-zero average $S$, and has novel consequences for direct detection experiments. First, the random spin population of VDM will result in a randomly distributed dark electric field, similar to the random linear polarization case~\cite{Arias:2012az}. Second, the circular polarization at the Compton time scale of the spin-aligned VDM population can make the dark electric field insensitive to the haloscope's maximum sensitivity, which is upper bounded by the sidereal rotation of the haloscope. Finally, the strength of the signal's sidereal modulation is given by the fraction of the spin-aligned VDM component, which can be used to investigate the composition of VDM in the scenario of signal detection.

It is possible to have still more general polarization settings using the correlation matrix formalism we presented in Ref.~\cite{PhysRevD.108.083021}. Such a setting can tune the desired value of $S^i$ in any component, and thus the value of $\kappa$ relevant for experiment (the case with very small $S/N$ is generated spontaneously by numerical effects in our CF simulations). Such a random mixed state model is generated spontaneously by our random initial conditions from the power spectrum. 

In future, we intend to perform fully cosmological simulations with more realistic initial conditions composed for specific particle physics and early Universe models of VDM, and thus make more accurate predictions on the spin polarization distribution for the growing experimental effort to search for VDM.

\appendix

\section*{Acknowledgements}
We thank Xiaolong Du for providing significant assistance during our research, and collaboration in an early stage of this work. We thank Lei Pan, Chao Gao, Yikun Gu, Hongyi Zhang, ZhiPan Li and Mudit Jain for beneficial discussions, and Mudit Jain for reading an early version of this manuscript.

JC is supported by Fundamental Research Funds for the Central Universities (SWU-KR22012) and Chongqing Natural Science Foundation General Project (2023NSCQ-MSX1929). JC and LHN acknowledge support by the Deutsche Forschungsgemeinschaft (DFG, German Research Foundation) under Germany’s Excellence Strategy – EXC 2121  \enquote{Quantum Universe} – 390833306. DJEM is supported by an Ernest Rutherford Fellowship from the STFC, Grant No. ST/T004037/1.

This research made use of computational resources at the School of Physical Science and Technology, Southwest University. JC especially appreciates Zhipan Li for his assistance in setting up the cluster and for the the Fundamental Research Funds for the Central Universities. 

\bibliography{reference}

\end{document}